# Lateral Pressure Profile in Bent Lipid Membranes and Mechanosensitive Channels Opening: Microscopic Theory

Anna A. Drozdova[1] and Sergei I. Mukhin[2]

Department of Theoretical Physics and Quantum Technologies,

Moscow institute for Steel and Alloys,

4 Leninsky prospekt, 119049, Moscow, Russia

**ABSTRACT**

This review describes the analytical calculation of lateral pressure profile in lipid membrane with finite curvature. The derivation is based on previously the microscopic model of flexible strings [1,2]. According to this theory the energy per unit chain is considered as energy of flexible string (Euler's elastic beam of finite thickness) and interaction between chains is considered as an entropic repulsion. This microscopic theory allows to obtain expression for lateral pressure distribution in bent lipid membrane if treating a bending as a small deviation from the flat membrane conformation and using perturbation theory with small parameter $L_0 J$, where $L_0$ is the monolayer thickness, $J$ is mean curvature of membrane. Because lateral pressure distribution is related to the elastic properties of lipid bilayer [3] then first spontaneous moment of lateral pressure and the expression for bending modulus may be derived from this theoretical model. Finally, based on analytical expression for the lateral pressure distribution in bent lipid membrane one can obtain analytically the energy difference between two states of mechanosensitive channel, particularly MscL [4, 5, 6]. Calculated expression depends on inter-facial surface tension, the channel area change in region between hydrophobic interior and the aqueous surroundings and the bending modulus and spontaneous bending moment values.

[1]contact e-mail: *anechka.drozdova@gmail.com*;  [2]contact e-mail: *i.m.sergei.m@gmail.com*

# I. INTRODUCTION

The relevance of the present study is supported by the fact that biological membranes constitute complex supramolecular structures that surround all living cells and form them in closed organelles. The basic structural element of the membrane is lipid bilayer, whose mechanical and thermodynamical properties are believed to play important role in the functioning of organized matter.

The lipids have one polar (hydrophilic) "head" and two non-polar (hydrophobic) "tails" (chains). There is a great variety of membrane constituents and lipid characteristics (head group type, chain length, etc). Our analytical theory of lipid membranes, based on the microscopic model, was developed in [1,2]. In this model each chain is modeled as a flexible string with finite bending rigidity and incompressible area per chain is found self-consistently. The ensemble of chains is investigated within a mean-field approximation with respect to the strings collisions. According to this approximation a steric (entropic) repulsion between the lipid tails is represented with an effective lateral potential acting on each string in each monolayer. The negative lateral tension keeping all the strings (i.e. lipids) together comes from hydrophobic-hydrophilic interface in the head-groups area of a membrane and is treated as an input parameter of the model. So, one can write the energy functional for a lipid chain and then find the partition function and free energy of the chains. The strength of the effective lateral potential acting on each string, that minimizes the free energy, is found then in a self-consistent (mean-field) manner. Knowing the free energy one can find various macroscopic characteristics of the lipid bilayer.

Our aim is to derive analytical expression for the lateral pressure profile in a bilayer membrane with finite curvature. Our theory is quite general, but for a comparison of our analytical results with experiments we use as the reference typical microscopic parameters of e.g. DPPC lipids: monolayer thickness $L_0 = 1,5\, nm$, incompressible area per chain $A_0 = 0,2\, nm^2$. The flexural rigidity of a lipid tail can be evaluated then from the Flory's polymer theory: $K_f \approx k_B T L_0 / 3$, $T = T_0$ where $T_0 = 300\, K$ is a reference temperature. The lateral pressure is a very important feature of lipid membrane because it is related to the elastic properties of lipid bilayer and also membrane protein functionality.

The article is organized as follows. In Sec. II we describe the basics of the flexible strings model and give a typical example of calculation of the lateral pressure profile in a flat bilayer. In Sec. III, we introduce mathematical description of the bent lipid bilayer via the depth coordinate dependent area per chain according to differential geometry relation for the stack of bent surfaces with finite curvature radius. In Sec. IV we find how the strength coefficient of entropic potential changes as function of the depth coordinate inside the membrane and calculate analytically the lateral pressure profile in lipid membrane with finite curvature using the perturbation theory for the small parameter $L_0 J$. Dependence of membrane protein functionality on lateral pressure profile are discussed in Sec. V. The results and their comparison with the other calculations of the thermodynamic characteristics of the bent membrane are presented in Sec. VI. Derivation of the analytical solution of the self-consistency equation, first and second derivatives of entropic coefficient with respect to area per chain, the analytical calculation of bending modulus via the lateral pressure moment, calculation details of the energy difference between two states of mechanosensitive channel and calculations details for various lipids are represented in Appendices A, B, C, D and F and E, respectively.

## II. FLEXIBLE STRINGS MODEL. ANALYTICAL DERIVATION OF THE LATERAL PRESSURE PROFILE

Microscopic model of lipid membrane as an ensemble of flexible strings was investigated in [1,2]. Below we describe the basic thesis of our theoretical model.

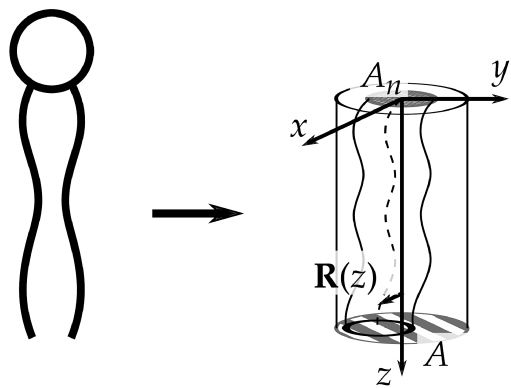

Figure 1 - Hydrocarbon tails of lipid as flexible strings: R (z) - the deviation of the string from the cylinder axis, $A_0$ is incompressible area per chain. The figure adapted from the article [1].

Lipid chain is considered as Euler's elastic beam of finite thickness (fig. 1), and its energy functional is written in the following form:

$$E_t = \int_0^L \left( \frac{\rho \dot{R}^2(z)}{2} + \frac{K_f}{2} \cdot \left( \frac{\partial^2 R(z)}{\partial z^2} \right)^2 + \frac{B}{2} \cdot R^2(z) \right) dz, \quad (1)$$

here $R(z)$ is the deviation of the beam from the straight line at each level $z$.

In formula (1) the first term is the kinetic energy of the chain, the second term is the bending energy of the chain, the third term is the potential representing entropic repulsion with the neighboring chains; $\rho$ is the linear density of the chain.

Energy functional of the chain conformations (1) can be rewritten, provided there are certain boundary conditions at the chain ends [1, 2], in the operator form:

$$E_t = \frac{1}{2} \cdot \int_0^L \left[ \rho \dot{R}^2(z) + R(z) \hat{H} R(z) \right] dz, \quad (2)$$

where the conformational energy operator is determined by expression:

$$\hat{H} = K_f \cdot \frac{\partial^4}{\partial z^4} + B \quad (3)$$

We can solve the operator equation $\hat{H} E_n = E_n \cdot R_n$ and find the eigenfunctions and eigen values of $\hat{H}$. Then we expand the chain deviation $R(z)$ in these eigenfunctions:

$R(z) = \sum_n C_n R_n(z)$. Using the orthonormality condition $\int_0^L R_n(z) R_{n'}(z) dz = \delta_{nn'}$, one can write the energy functional in the amplitudes $\{C_n\}$ representation of the expansion over the eigenfunctions:

$$E_t = \frac{1}{2} \cdot \sum_n \left[ \rho \dot{C}_n^2 + C_n^2 E_n \right], \quad (4)$$

where $E_n$ is the n-th eigenvalue:

$$E_n = B + \frac{K_f}{L^4} \cdot \left( \frac{\pi}{4} \right)^4 (4n-1)^4; n \geq 1; E_0 = B \quad (5)$$

of the n-th eigenfunction $R_n(z)$:

$$R_n(z) = q_n \cdot \left[ \cos(\lambda_n z) + \frac{\cos(\lambda_n L)}{\cosh(\lambda_n L)} \cdot \cosh(\lambda_n z) \right]; n \geq 1; R_0(z) = \sqrt{\frac{1}{L}} \quad (6)$$

In last formula we use the following notations:

$$q_n \approx \sqrt{\frac{2}{L}} \quad (7)$$

and

$$\lambda_{n \geq 1} \cdot L = -\frac{\pi}{4} + \pi n \quad (8)$$

Using the energy functional (1), we find the partition function:

$$Z = \int \exp\left(-\frac{E_t}{kT}\right) DR\ D\dot{R} \qquad (9)$$

The partition function in the amplitudes $\{C_n\}$ representation takes the form:

$$Z = \prod_n \int \exp\left(-\frac{p_n^2}{2\rho k_B T} - \frac{C_n^2 \cdot E_n}{2 k_B T}\right) dp_n dC_n = \prod_{n\geq 0} \frac{k_B T}{\hbar} \cdot \sqrt{\frac{\rho}{E_n}} = \prod_{n\geq 0} \frac{k_B T}{\hbar \omega} \qquad (10)$$

Hence, using the free energy formula:

$$F_t = -k_B T \ln Z, \qquad (11)$$

where $F_t$ is free energy of the tails, and then substituting (10) into (11) one can obtain:

$$F_t = k_B T \sum_n \ln E_n + const, \qquad (12)$$

Using formula for the free energy (12) we find the lateral pressure inside the hydrophobic part of the bilayer:

$$P_t = -\frac{\partial F_t}{\partial A}, \qquad (13)$$

where $P_t$ is the lateral pressure produced by hydrophobic chains.

The self-assembly condition (i.e. zero total lateral tension of the self-assembled membrane) is represented in the form:

$$\left(\frac{\partial F_t}{\partial A}\right) + \gamma = 0, \qquad (14)$$

where $\gamma$ is the surface tension at the interface between the polar and hydrophobic regions of the membrane [1]. The physical meaning of this value is the surface tension at the lipid-water interface (situated between hydrophobic tails and hydrophilic heads): $\gamma = 30-100\ dyn/cm$ for various lipids. So the equilibrium condition (14) simply states that repulsion between chains should be balanced by the surface tension $\gamma$.

Lateral pressure is expressed in the following form [1]:

$$P_t = -\int k_B T \cdot \sum_{n=0} \frac{R_n^2(z)}{E_n} \cdot \left(\frac{\partial B}{\partial A}\right)_T dz \equiv \int \Pi_t(z) dz \qquad (15)$$

Here the lateral pressure profile distribution $\Pi_t(z)$ for the flat bilayer is introduced by the formula:

$$\Pi_t^{(0)}(z) = -kT \sum_n \frac{R_n^{(0)2}(z)}{E_n^{(0)}} \frac{\partial B^{(0)}}{\partial A} \qquad (16)$$

# III. MODEL OF BENT BILAYER MEMBRANE

Here we introduce our description of bent bilayer. We consider the pure bending of the membrane, i.e. neutral surfaces of both monolayaers of the bilayer are not stretched [7,8] (see fig. 2). We assume that the neutral surfaces are located in the hydrophobic-hydrophilic interfaces of the bilayer membrane [9,10]. Conservation of volume per lipid causes top layer to become thicker while the bottom one becomes thinner. Below we use notations $A_b(z)$ for the area per chain in the inner ('bottom') monolayer, and $A_t(z)$ for the outer ('top') monolayer respectively (see fig. 2).

As is well known from differential geometry [8], for two parallel curved surfaces at a distance $z$, the area elements are related by the following equation: $dA'=dA\cdot(1-zJ+z^2 K)$, where $J$ is mean curvature; $K$ is Gaussian curvature. In what follows we disregard Gaussian curvature contribution for the chosen topology of the membrane and write the area change in the form [11]:

$$dA'=dA\cdot(1-zJ), \qquad (17)$$

where $J$ is the curvature of the surface $dA$.

Using formula (17), one may define the area change via depth coordinate $z$ for the 'bottom' monolayer in a curved bilayer membrane (the mean curvature of the bottom monolayer is represented by expression: $J_b=-J(1+L_b J)<0$ ) and for the 'top' monolayer ( $J_t=J(1-L_t J)>0$ ):

$$\begin{array}{l} A(z)=A(1-zJ+z\,L_t J^2);\, 0\leq z \leq L_t\,; top \\ A(z)=A(1+zJ+z\,L_b J^2);\, 0\leq z \leq L_b\,; bottom \end{array} \qquad (18)$$

In eq. (18) the origin of $z$ is different for 'bottom' and 'top' monolayer (see Fig. 2), $A$ - the area per lipid chain in flat monolayer ( $R=\infty \to J=0$ ) and $A=A_0\cdot a$, where $A_0$ is incompressible area per lipid chain and $a$ is dimensionless area per lipid chain (because the lipid strings makes the bending fluctuations its real area is greater than incompressible one - see App. A).

Using formula (18) one can define the derivatives of the area per lipid chain with respect to curvature $J$ in each monolayer:

$$\begin{array}{l} \dfrac{\partial A}{\partial J}(J=0)=-A\,z\,; top \\ \dfrac{\partial A}{\partial J}(J=0)=A\,z\,; bottom \end{array} \qquad (19)$$

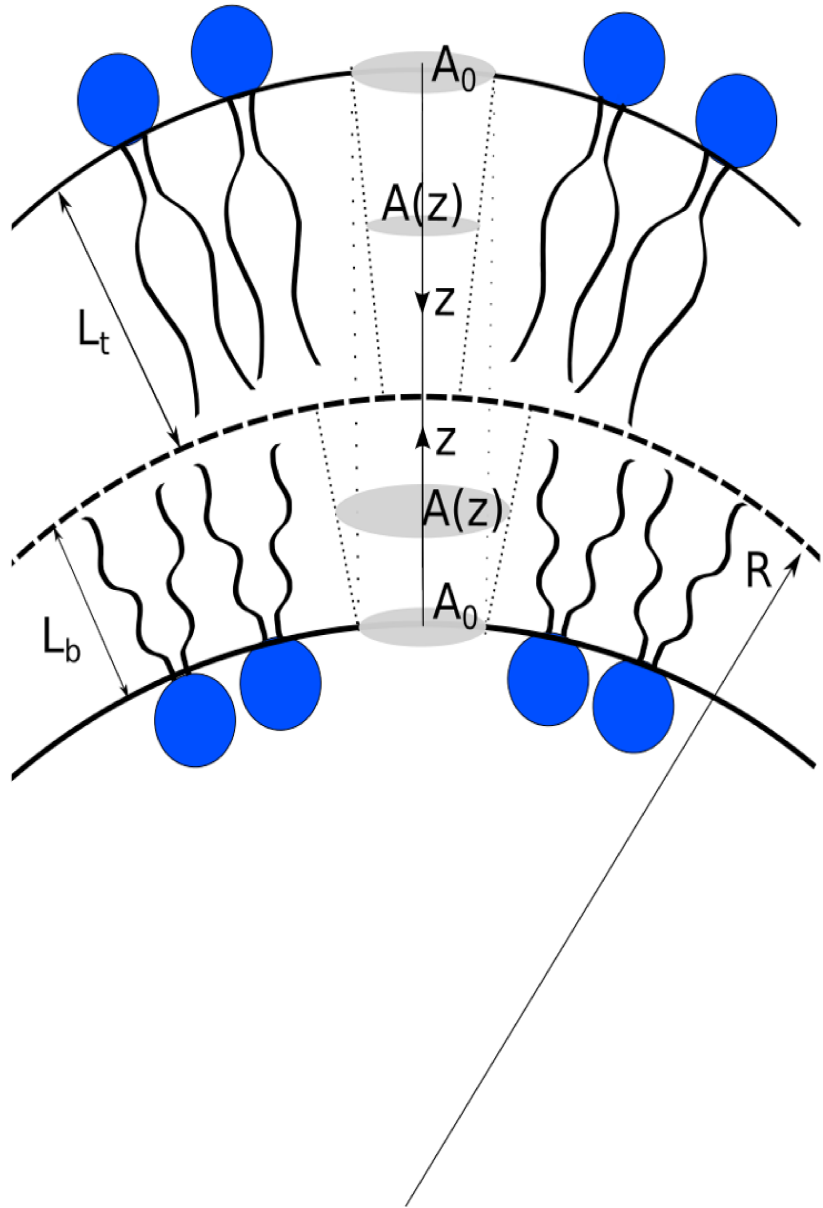

Figure 2 – Membrane bending model: $L_b$ - the thickness of the inner monolayer, $L_t$ - the thickness of the outer monolayer; $|J_b|=\dfrac{1}{R_b}$, $|J_t|=\dfrac{1}{R_t}$ - the mean curvature of bottom and top cylindrical neutral surfaces, respectively, where $R_b=R-L_b$ and $R_t=R+L_t$ - the curvature

radii of these surfaces. Note that for spherical surface the mean curvature for monolayers are: $|J_b|=\dfrac{2}{R_b}$, $|J_t|=\dfrac{2}{R_t}$. $J$ is curvature of the intermonolayer surface. The curvatures of bottom and top neutral surfaces differ by sign. Under bending area per lipid at the headgroup neutral surfaces of each monolayer is unchanged with respect to flat membrane provided there is a lipids reservoir or intermonolayer slide.

## IV. THE CHANGE OF LATERAL PRESSURE PROFILE INDUCED BY CURVATURE

In this section we re-derive in the present context the formula (16). Namely, one can apply perturbation theory to the conformational energy operator $\hat{H}$ (3) and find analytically the first correction to the lateral pressure profile due curvature. The unperturbed 'Hamiltonian' for flat bilayer is given by the formula:

$$\hat{H}^{(0)} = K_f \cdot \dfrac{\partial^4}{\partial z^4} + B^{(0)} \tag{20}$$

In the case of bent membrane the entropic repulsion coefficient $B^{(0)}$ of the flat bilyer in formula (20) acquires the curvature induced depth-dependent correction $B^{(1)}(z,J)$. Thus, the perturbation of 'Hamiltonian' is $\hat{H}^{(1)}$:

$$\hat{H}^{(1)} = B^{(1)}(z,J) \tag{21}$$

Further function $B^{(1)}(z)$ constituting the perturbation operator in (21) can be obtained in the first approximation in $L_0 \cdot J$ by the Taylor's expansion of $B$ in powers of the curvature $J$:

$$B = B^{(0)} + \dfrac{\partial B}{\partial J} \cdot J + \dfrac{1}{2} \dfrac{\partial^2 B}{\partial J^2} \cdot J^2 \tag{22}$$

In this case derivatives of the entropic potential $\dfrac{\partial B}{\partial J}$ and $\dfrac{\partial^2 B}{\partial J^2}$ are taken at zero curvature $J=0$.

The first correction to entropic potential:

$$B^{(1)} = \dfrac{\partial B}{\partial J} \cdot J, \tag{23}$$

where derivative of $B$ with respect to curvature $\dfrac{\partial B}{\partial J}$ equals:

$$\dfrac{\partial B}{\partial J} = \dfrac{\partial B^{(0)}}{\partial A} \dfrac{\partial A}{\partial J}(J=0) \tag{24}$$

We apply perturbation theory $(\hat{H}^{(0)} + \hat{H}^{(1)}) R_n = E_n R_n$ to obtain the first correction $E_n^{(1)}$ to

eigenvalue $E_n^{(0)}$:

$$E_n^{(1)} \equiv B_{nn} = \int_0^{L_0} B^{(1)}(z,J) \cdot R_n^{(0)2}(z) dz \qquad (25)$$

Then we can re-write the free energy for curved membrane in the operator form:

$$F^{curv} = kT \sum_n \ln(E_n^{(0)} + E_n^{(1)}), \qquad (26)$$

Differentiating with respect to area per chain *A(z)*, we can find lateral pressure profile in curved membrane $\Pi_t^{curv}(z,J) = -\frac{\delta F^{curv}}{\delta A(z)}$:

$$\Pi_t^{curv}(z,J) = -kT \cdot \sum_n \frac{1}{E_n^{(0)} + E_n^{(1)}} \cdot \frac{\delta E_n}{\delta B(z)} \cdot \frac{\partial (B^{(0)} + B^{(1)})}{\partial A} \qquad (27)$$

Using formula (25) one can calculate the variational derivative $\frac{\delta E_n}{\delta B(z)}$ and substituted this expression in (27) obtain:

$$\Pi_t^{curv}(z,J) = -kT \cdot \sum_n \frac{R_n^{(0)2}(z)}{E_n^{(0)}} (1 - \frac{E_n^{(1)}}{E_n^{(0)}})(\frac{\partial B^{(0)}}{\partial A} + \frac{\partial B^{(1)}}{\partial A}) \qquad (28)$$

In last formula we may distinguish the lateral pressure profile in flat membrane $\Pi_t^{(0)}(z)$ and the correction to the lateral pressure profile due curvature $\Pi_t^{(1)}(z,J)$:

$$\Pi_t^{(1)}(z,J) = -kT \sum_n \frac{R_n^{(0)2}}{E_n^{(0)}} \cdot \frac{\partial B^{(1)}}{\partial A} + kT \sum_n \frac{R_n^{(0)2}}{E_n^{(0)}} \cdot \frac{E_n^{(1)}}{E_n^{(0)}} \cdot \frac{\partial B^{(0)}}{\partial A} + kT \sum_n \frac{R_n^{(0)2}}{E_n^{(0)}} \cdot \frac{E_n^{(1)}}{E_n^{(0)}} \cdot \frac{\partial B^{(1)}}{\partial A} \qquad (29)$$

As one can see from equations (23), (24) and (25) the last term in formula (29) is proportional to the second order curvature $J^2$ and it might be vanished in first approximation. So, the final expression to the correction to lateral pressure due curvature is represented by formula:

$$\Pi_t^{(1)}(z,J) = -kT \sum_n \frac{R_n^{(0)2}(z)}{E_n^{(0)}} \cdot \frac{\partial B^{(1)}}{\partial A} + kT \sum_n \frac{R_n^{(0)2}(z)}{E_n^{(0)}} \cdot \frac{E_n^{(1)}}{E_n^{(0)}} \cdot \frac{\partial B^{(0)}}{\partial A}, \qquad (30)$$

where $E_n^{(1)}$ and $\frac{\partial B^{(1)}}{\partial A}$ can be re-written as:

$$E_n^{(1)} = J \cdot \frac{\partial B^{(0)}}{\partial A} \cdot \int_0^{L_0} \frac{\partial A}{\partial J} R_n^{(0)2}(z) dz \qquad (31)$$

$$\frac{\partial B^{(1)}}{\partial A} = J \cdot \frac{\partial^2 B^{(0)}}{\partial A^2} \frac{\partial A}{\partial J} \qquad (32)$$

Derivatives of $B^{(0)}$ with respect to $A$ are derived in Appendix B. One may re-write formula (30) using the expressions for area per chain change (18) and obtain the corrections to the lateral pressure profile $\Pi_t^{(1)}(z)$ in inner and outer monolayers of bent bilayer:

$$\Pi_{t,bot}^{(1)}(z) = -kT \frac{\partial^2 B^{(0)}}{\partial A^2} \cdot A \cdot \sum_n \frac{R_n^{(0)2}(z)}{E_n^{(0)}} \cdot z J + kT \left( \frac{\partial B^{(0)}}{\partial A} \right)^2 \cdot A \sum_n \frac{R_n^{(0)2}(z)}{E_n^{(0)2}} \cdot J \int_0^{L_0} z R_n^{(0)2} dz, \qquad (33)$$

where $A = a \cdot A_0$ - the area per lipid chain in flat bilayer.

In expression (33) variable $z$ varies in the interval $z \in [0; L_b]$, where $L_b = L_0 - \frac{L_0^2 J}{2}$ is the thickness of the inner ('bottom') monolayer that follows from the volume per lipid conservation, see fig. (2).

$$\Pi_{t,top}^{(1)}(z) = kT \frac{\partial^2 B^{(0)}}{\partial A^2} \cdot A \cdot \sum_n \frac{R_n^{(0)2}(z)}{E_n^{(0)}} \cdot z J - kT \left( \frac{\partial B^{(0)}}{\partial A} \right)^2 \cdot A \sum_n \frac{R_n^{(0)2}(z)}{E_n^{(0)2}} \cdot J \int_0^{L_0} z R_n^{(0)2} dz, \qquad (34)$$

In expression (34) variable $z$ varies in the interval: $z \in [0; L_t]$, where $L_t = L_0 + \frac{L_0^2 J}{2}$ is the thickness for the outer ('top') monolayer, derived from conservation condition of the volume per lipid (here $J$ is curvature of the intermonolayer surface – see fig. 2)

The results (33)-(34) are plotted in Figure 3.

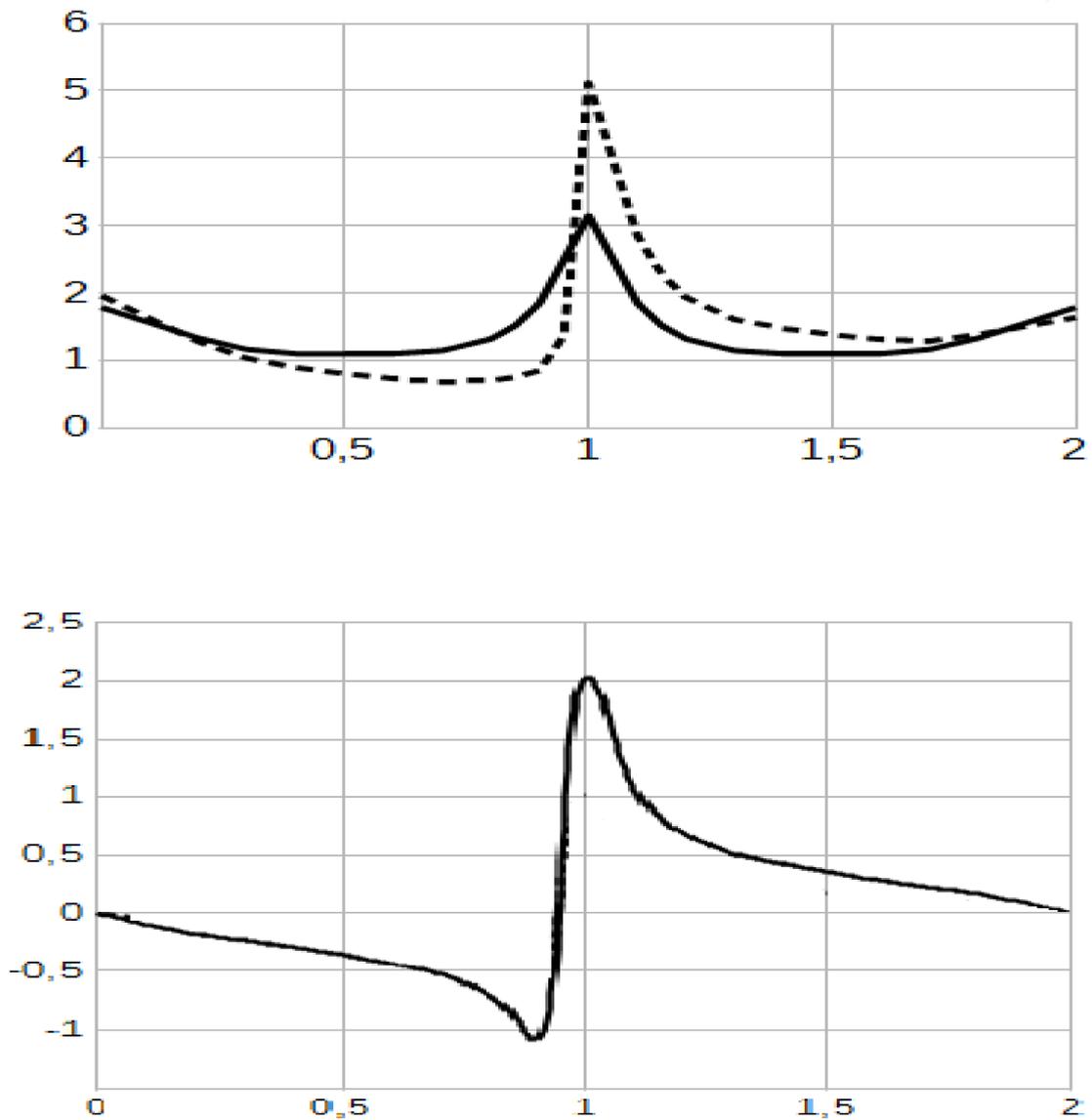

Figure 3 – 3A: Lateral pressure profiles in hydrophobic part of DPPC bilayer, normalized with $\Pi_0 = \gamma/L_0$, for the flat membrane (solid line), lateral pressure profile in lipid bilayer with finite curvature $J = \frac{2}{R}$ (dotted line), where $R = 20\,nm$ - fixed curvature radius. Here $\gamma = 50\,dyn/cm$ is interfacial surface tension; 3B: the difference between the lateral pressure profiles in the flat and in the bent bilayer with curvature radius $R = 20\,nm$ (solid line).

As one can note the bending shifts position of the peak of lateral pressure profile along the z-axis. This is caused by the fact that the top and bottom monolayers acquire different thicknesses due to the bending (see Fig. 2).

Inside the inner $z \in [0; L_b]$ of the bent bilayer (see Fig. 2) the area per one tail is greater than in the flat membrane (area change is $\delta A_b(z)$ positive - see eq. (18)), therefore, the entropic repulsion is weakened, and, hence, the lateral pressure is lowered (see Fig. 3A). In the outer layer $z \in [0; L_t]$ of bilayer inter-tails space is compressed (the area change is $\delta A_t(z)$ negative) and entropic repulsion between tails, i.e. lateral pressure, raises up (see Fig. 3A).

## V. MECHANOSENSITIVE CHANNEL GATING

Mechanosensivity is a general sensory mechanism found in living organisms. The simplest cell structures, for example, the lipid bilayer vesicle, can respond to mechanical deformation by elastic membrane dilation and curvature changes. It's known that the functionality of many membrane proteins, particularly MscS, MscK, bacterial membrane protein MscL (the most studied), is sensitive to the lipid environment [12, 13, 14] and form mechanically gated channels when they embedded in lipid bilayer. In each organism MSCs perform the certain functions. For people, for example, MSCs are responsible for the regulation of blood pressure and play a role of the pain receptors [15].

Because mechanosensitive channels (MSCs) are inserted in a lipid bilayer, they are sensitive to its deformations [16]. MSCs respond to mechanical stress by changing of the channel shape between the closed and open state. The change in external dimensions between the closed and open states associated with the opening of a pore through which anions and cations can pass [14], [17]. The conformation shape of MscL was found to change between cylindrical (closed) and cone (open) shapes [18]. Other authors [19] assumed that cone shape of channel corresponds to its intermediate state, but hour-glass shape corresponds to its open state (see fig. 4).

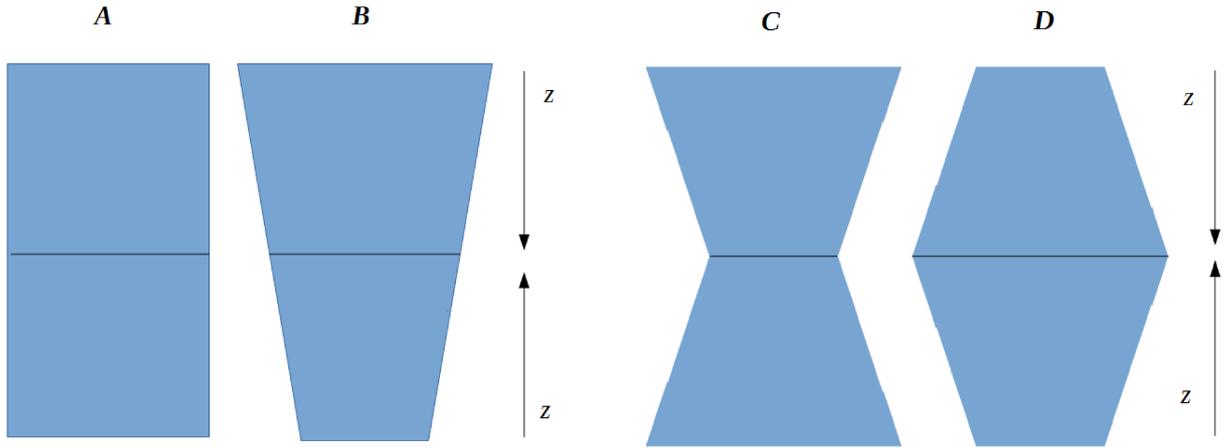

Figure 4 - The shape of the mechanosensitive channel; 4A: the cylindrical channel; 4B: conical model of the channel; 4C, D: the hour-glass shape of the channel; the area in the middle of MSC doesn't change – see below eq. (36)

To estimate the effect of pressure profile on membrane proteins, we followed the approach introduced in article [4], [20], [21] where was shown that lateral pressure profile affects the functioning of the proteins embedded in lipid membrane and was calculated numerically the work on MscL opening:

$$\Delta E = \int_0^{2L_0} P(z) \Delta A(z) dz, \qquad (35)$$

where $P(z)$ – lateral pressure in hydrophobic part of the bilayer, $\Delta A(z)$ — the channel area changing during opening.

In article [17] was experimentally determined (by the electron paramagnetic resonance methods) that mechanosensitive channel opening of pore with a diameter $d = 2,5 \, nm$.

The authors of work [5] using the molecular dynamic methods found that lipid bilayer bending is the cause why MSCs can to be opened. Our aim in this section is to test analytically this hypothesis and to develop our previously started calculation [22] and obtain the expression for energy difference between two different states of mechanosensitive channel.

At the beginning we define the area difference between protein states using experimental data [17]:

$$\begin{aligned} A(z) &= A_m + \Delta A(z) = A_m \pm A_m \cdot \alpha z \,, conical \\ A(z) &= A_m + \Delta A(z) = A_m + A_m \cdot \alpha z \,, hour-glass \end{aligned} \qquad (36)$$

where $A_m = \frac{\pi d^2}{4}$ is changing of the MSC area in the middle of bilayer and the slope is: $\alpha = \frac{3}{10 L_0}$, where $L_0$ is the monolayer thickness. Here origin of axis $z$ begins in the center of the bilayer [17].

Then for the conical shape of the mechanosensitive channel (fig. 4B) we obtain the next area changing:

$$\Delta A_{A \to B}(z) = \alpha A_m(z-L); bot$$
$$\Delta A_{A \to B}(z) = -\alpha A_m(z-L); top \tag{37}$$

also for hour-glass conformation of the channel (fig. 4C):

$$\Delta A_{A \to C}(z) = -\alpha A_m(z-L); bot$$
$$\Delta A_{A \to C}(z) = -\alpha A_m(z-L); top \tag{38}$$

and for hour-glass conformation of the channel (fig. 4D), respectively:

$$\Delta A_{A \to D}(z) = \alpha A_m(z-L); bot$$
$$\Delta A_{A \to D}(z) = \alpha A_m(z-L); top \tag{39}$$

In eq. (37)-(39) zero of axis $z$ is in neutral surfaces (see fig. 1) for calculation in our microscopic theory.

Membrane may be bend bend spontaneously with curvature $J_s$ or under the influence of thermal fluctuations and its surface may become concave or convex shape with finite curvature $J$.

We can write the energy difference between two different states of channel in the following form [23]:

$$\Delta W = W_{final} - W_{initial}, \tag{40}$$

where $W_{final}$ and $W_{initial}$ - the energy of final and initial state of lipid environment and the channel.

The comparison of calculations results for the energy difference between two states of mechanosensitive channel in conical and hour-glass models are represented in Tables 1 and 2 (for more details one can view App. C, D, E, F).

Decreased opening barrier due lateral pressure imbalance were located separately in figures 5, 6. So, one can conclude that when MSC changed its shape from cylindrical to cone, the bending energy of bilayer decreases by $5 k_B T$ average. During this process the stretching energy is increasing by amount about $10 k_B T$. These estimate agrees qualitatively well with results of molecular dynamics simulations presented in [4, 23, 24] where values of the energy difference between open and closed protein states is about $0-10 k_B T$. And when the channel has changed its shape from cylindrical to hour-glass one, the membrane hydrophobic energy increases even more, about $15-30 k_B T$, but the spontaneous bending energy reduces this contribution by amount

$8-15\,k_B T$ .

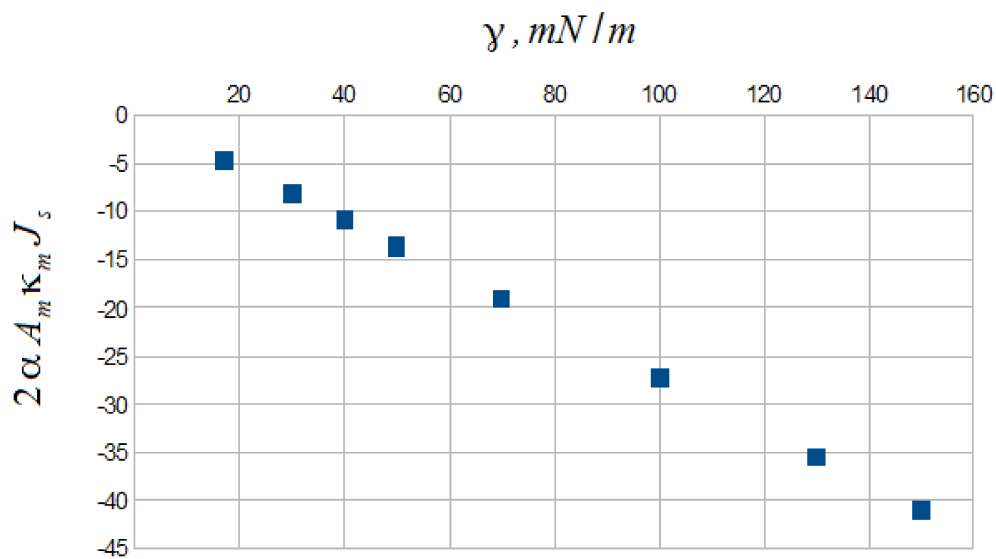

Figure 5 – The influence of spontaneous bending of lipid membrane on values of the energy difference between cylindrical and hour-glass conformations of the mechanosensitive channel on DPPC bilayer environment with monolayer thickness $L_0 = 1,5\,nm$ , incompressible area per chain $A_0 = 0,2\,nm^2$ and temperature $T = 300\,K$ . Here $\gamma$ - the surface tension between polar and hydrophobic regions. One can prove (App. F) that for any monolayer thickness will be obtained the almost constant value of energy difference provided the mechanosensitive channel embedded into the flat membrane.

| Process details | $\Delta W_{A \to B}$ | $\Delta W_{A \to C}$ | $\Delta W_{A \to D}$ |
|---|---|---|---|
| $\Pi_f = \Pi_i + \Pi^{(1)}$, $A_f = A_i \equiv A_m + \Delta A(z)$, $\Delta W = \int_0^{2L_0} \Pi^{(1)}(z,J) \Delta A(z) dz$ | $-2\alpha A_m \kappa_m J + \Delta A_{max} \widetilde{K}_{Ab} L_0 J$ | 0 | 0 |
| $\Pi_f = \Pi_i \equiv \Pi^{(0)}$, $A_f = A_i + \Delta A(z) \equiv A_m + \Delta A(z)$, $\Delta W = \int_0^{2L_0} \Pi_t^{(0)}(z) \Delta A(z) dz$ | 0 | $2\alpha A_m \kappa_m J_s + 2\Delta A_{max} \gamma$ | $-2\alpha A_m \kappa_m J_s - 2\Delta A_{max} \gamma$ |
| $\Pi_f = \Pi_i + \Pi^{(1)}$, $A_f = A_i + \Delta A(z) \equiv A_m + \Delta A(z)$, $\Delta W = \int_0^{2L_0} \Pi_t^{(0)}(z) \Delta A(z) dz + \int_0^{2L_0} \Pi^{(1)}(z,J) \Delta A(z) dz$ | $-2\alpha A_m \kappa_m J + \Delta A_{max} \widetilde{K}_{Ab} L_0 J$ | $2\alpha A_m \kappa_m J_s + 2\Delta A_{max} \gamma$ | $-2\alpha A_m \kappa_m J_s - 2\Delta A_{max} \gamma$ |

Table 1 – The energy difference between two states of the mechanosensitive channel (see formula (40)) in two models: conical and hour-glass, here $\kappa_m$ - the bending modulus of the monolayer (App. C), $\kappa_m J_s$ - the first spontaneous bending moment (App. C), $L_0$ - the monolayer thickness, $J$ - the curvature of intermonolayer surface (fig. 2), $A_0$ - the incompressible area per lipid chain, $a,b$ - the dimensionless area per lipid chain and the entropic repulsion coefficient, respectively, that found self-consistently [1] (App. A), $A_m = 4,9\, nm^2$ - the area of the MSC (formula (36)). The last line in this table shows the mixed process in which at first step the spontaneous curvature was induced in one of the monolayers of bilayer and the this bilayer was bent in addition. Here notification $i$ means the initial state, and notification $f$ means the final state of the mechanosensitive channel and its environment (lipid bilayer), $\Delta A_{max} = \alpha L_0 A_m \equiv \frac{3}{10} A_m$ - the channel area changing at inter-facial region (eq. (37)-(39)), $\widetilde{K}_{Ab} = 2\frac{kT}{A_0} a [\frac{\partial^2 b}{\partial a^2} \sum_{n=0}^{\infty} \frac{I_1(n)}{b+c_n^4} - (\frac{\partial b}{\partial a})^2 \sum_{n=0}^{\infty} \frac{I_1(n)}{(b+c_n^4)^2}] = r \cdot K_{Ab} \approx Const \cdot K_{Ab}$ - re-normalized area stretching modulus for lipid bilayer, where expression for integral $I_1(n)$ and $K_{Ab}$ one can find in App. C.

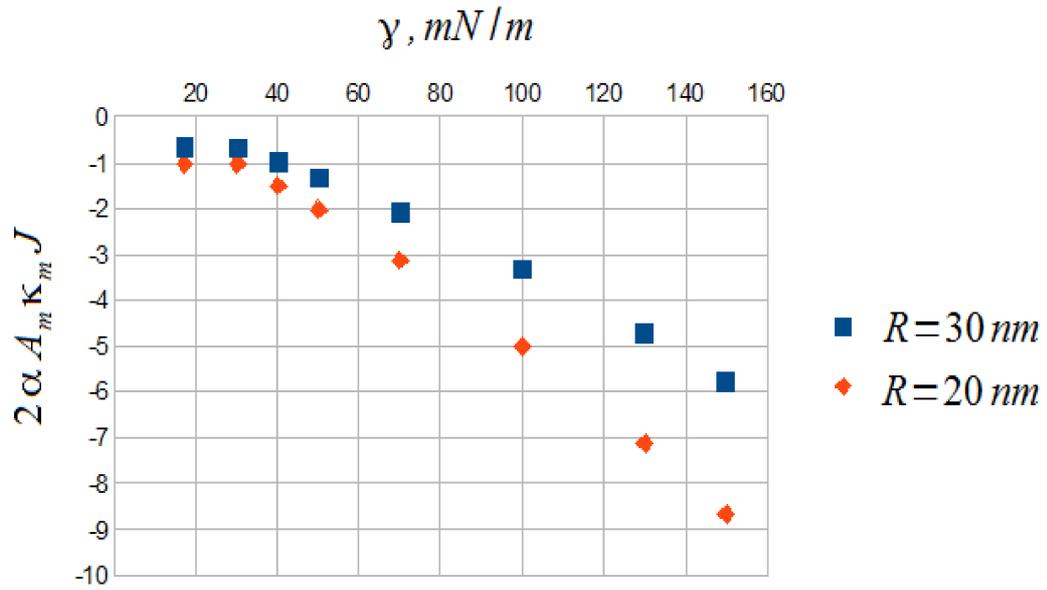

Figure 6 – The lateral pressure imbalance induced by curvature - so named the first moment of the bent lateral pressure profile in conical model of mechanosensitive channel, where $\gamma$ - the surface tension between polar and hydrophobic regions and $R$ – curvature radii. Here was the curvature for cylindrical surface $J = \dfrac{1}{R}$.

| Contribution into energy difference | $\Delta W_{A \to B}, k_B T$ | | | $\Delta W_{A \to C}, k_B T$ | | | $\Delta W_{B \to C}, k_B T$ | | |
|---|---|---|---|---|---|---|---|---|---|
| $\gamma, mN/m$ | 30 | 40 | 50 | 30 | 40 | 50 | 30 | 40 | 50 |
| $\lvert 2\alpha A_m \kappa_m J_s \rvert$ | 0 | 0 | 0 | -8,15 | -10,85 | -13,56 | -8,15 | -10,85 | -13,56 |
| $\lvert 2\Delta A_{max} \gamma \rvert$ | 0 | 0 | 0 | 21,33 | 28,44 | 35,55 | 21,33 | 28,44 | 35,55 |
| $\lvert 2\alpha A_m \kappa_m J \rvert$ | -2,08 | -3,02 | -4,04 | 0 | 0 | 0 | 2,08 | 3,02 | 4,04 |
| $\lvert \Delta A_{max} \tilde{K}_{Ab} L_0 J \rvert$ | 2,42 | 3,52 | 4,72 | 0 | 0 | 0 | -2,42 | -3,52 | -4,72 |
| $\Delta W_\Sigma$ | 0,34 | 0,5 | 0,68 | 13,18 | 17,59 | 21,99 | 12,84 | 17,09 | 21,31 |

Table 2 – The calculations of the energy difference between two conformations of MSC in DPPC bilayer environment ($A_0 = 0,2\ nm^2$ incompressible area per lipid chain, temperature $T = 300\ K$, monolayer thickness $L_0 = 1,5\ nm$) in conical and hour-glass model, fig. 4C. The result for hour-glass model, fig. 4D is different sign in comparison with hour-glass, fig. 4C (table 1). For calculation the first bending moment was taken the curvature radius $R=20\,nm$ and curvature for spherical surface $J=\frac{2}{R}$. Some calculation details are shown in App. E, F. As one can note that transitions between channel shapes are characterized by the relation between energy difference: $\Delta W_{A\to B\to C}=\Delta W_{A\to B}+\Delta W_{B\to C}\equiv\Delta W_{A\to C}$.

## VI. DISCUSSION

In this section we compare with other study our analytic calculation.

Obtained from microscopic flexible string model the lateral pressure profile in hydrophobic part of the bilayer (fig. 3) agrees qualitatively well with results of molecular dynamic simulation presented in [25] (fig. 7) and our previous obtained calculated pressure profile [26].

In other work [27] authors using molecular dynamic simulation investigated the lateral pressure profile asymmetry in the bent DOPC bilayer, with the bending due to asymmetric addition of non-bilayer lipids into one of the monolayer (fig. 8). Also authors [27] and [6, 28] investigated that asymmetric incorporation of conical lipids into bilayer was shown to activate mechanosensitive channels (particularly, the simplest model system such a bacterial MscL channel) in the absence of external surface tension.

Many authors discuss the activities in membrane proteins via perturbing the properties of lipid membrane, but general mechanism is not precisely understood. In work [29] authors comment and estimate the influence of certain effects, acting on mechanosensitive channels embedded in lipid membrane and they gave its own answer to the question what is actually sensed? And rather, MscL senses the tension in the membrane, namely the changes within lateral pressure profile [14, 16, 29, 30]. Authors of these studies concluded that some mechanosensors appear to be activated by amphipaths. But, the changes within lateral pressure profile rather than the curving of membrane in fact, are the stimulus for MscL gating [30] and also that the hydrophobic mismatch is not a primary stimulus for gating that was confirmed in [23].

As one can see our analytic calculation of the lateral pressure profile allows to estimate analytically the influence of the lateral pressure imbalance on the MSC gating mechanism and conclude that it has changed the MSC energy barrier $A_m\int_0^{2L_0}\Pi_t^{(0)}(z)dz=2\gamma A_m$ (tables 1, 2). And if changes in lipid composition induces spontaneous curvature, then these changes would be produce a significant conformational energy change. Also one can note that our calculation results allows to draw conclusion that the great tension $\gamma$ in hydrophobic region corresponds to a greater change in the energy barrier, because the bending modulus $\kappa_m$ and the entropic potential $b$ are increased with increasing $\gamma$. General result is represented in App. F. As one can see from table 9, mechanosensitive channels (and other ion channels) have been shown to be sensitive to inter-facial surface tension and spontaneous bending moment values. These results are well consistent with numerical calculations of previous study [4, 23]. Of course, one may conclude that bending of membrane associated with simple cylinder-cone-like shape channel transition doesn't produce a great changing in energy difference (tables 2, 9).

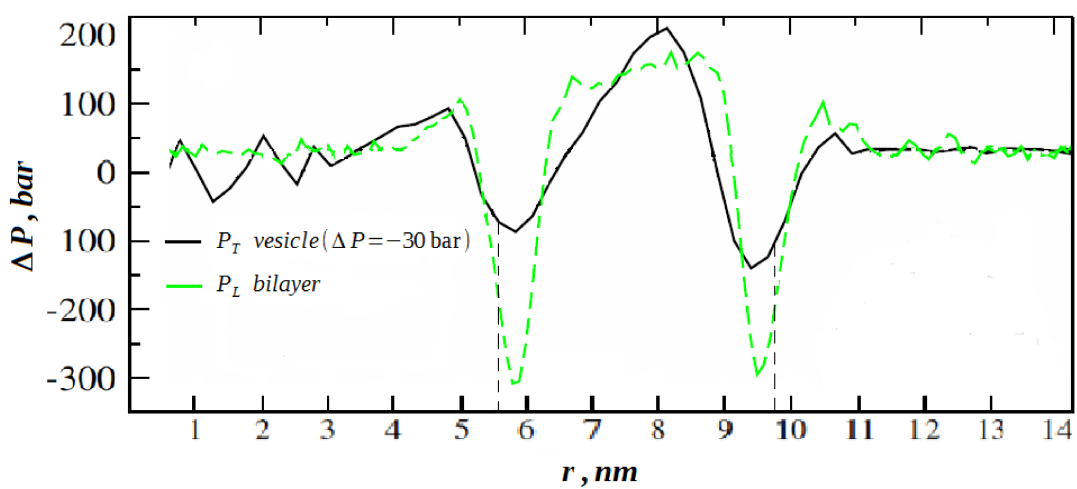

Figure 7 – Pressure profile for DOPC bilayer and vesicle, obtained by the molecular dynamic methods. Dotted vertical lines indicate the region corresponding to hydrophobic part of the bilayer. Asymmetric pressure profile in vesicle is shown with black solid line ( $\sigma = \frac{1}{2} \Delta P \cdot R$ - tension within a membrane, $\Delta P$ - the pressure across the membrane, $R$ - the curvature radius), for planar bilayer – with the green dotted line. In notation of article [25] these are $P_T$ and $P_L$, respectively. The figure adapted from article [25].

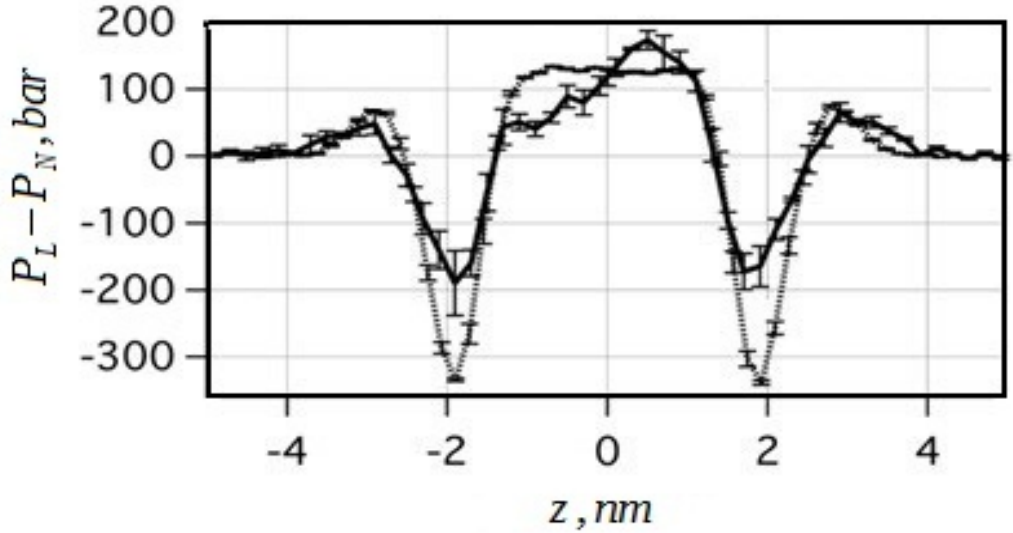

Figure 8 – Asymmetric addition of amphipaths in DOPC bilayer produces spontaneous curvature and the imbalance in lateral pressure profile – black solid line. Here gray dashed line indicates lateral pressure profile in the flat DOPC bilayer. The figure is adapted from article [27].

**ACKNOWLEDGMENTS**

The authors gratefully acknowledge the financial support of the Ministry of Education and Science of the Russian Federation in the framework of Increase Competitiveness Program of MISiS and the Russian Foundation for Basic Research (project no.130440327N (KOMFI)).

**APPENDIX A: SELF-CONSISTENCY EQUATION. ANALYTICAL DERIVATION**

In this section we derive the self-consistency equation analytically.

As one can see from (12) the partial derivative of the free energy $F$ with respect to entropic coefficient $B$ is:

$$\frac{\partial F}{\partial B} = k_B T \cdot \sum_n \frac{1}{E_n} \qquad (41)$$

On the other hand, from the initial expression for the energy functional of the chain (1) the following formula for partial derivative with respect to $B$ is expressed as follows:

$$\frac{\partial F}{\partial B} = k_B T \cdot \frac{\int\limits_0^L \int R_x^2(z) \exp\left[-E\frac{(R_x(z))}{k_B T}\right]}{\int \exp\left[-E\frac{(R_x(z))}{k_B T}\right]} = L \cdot \langle R^2 \rangle \qquad (42)$$

Here $\langle R^2 \rangle$ is the average of square deviation (fig. 9)

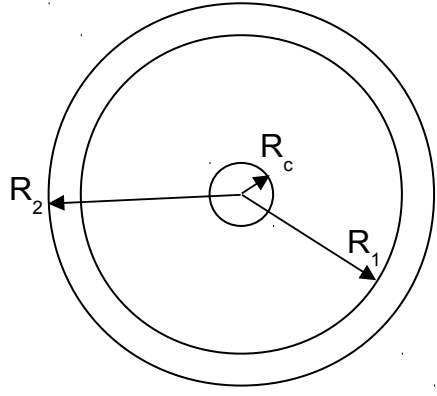

Figure 9 – The average deviation of the chain from the straight axis.
The figure adapted from article [1].

The average square deviation of the chain can be written in the form [1]:

$$\langle R^2 \rangle = \frac{\left(\sqrt{A} - \sqrt{A_0}\right)^2}{\pi}, \quad A = a \cdot A_0 \tag{43}$$

where $A_0$ is incompressible area per lipid chain.

Equating the right sides of the Eqs. (41) and (42), we obtain the self-consistency equation:

$$k_B T \cdot \sum_n \frac{1}{E_n} = \frac{L}{\pi} \cdot \langle R^2 \rangle \tag{44}$$

One can substitute the expression for the eigen values (5) into last formula (44) and obtain the self-consistency equation in dimensionless form [31]:

$$\frac{1}{b} + \sum_{n=1}^{\infty} \frac{1}{b + \left(\pi n - \frac{\pi}{4}\right)^4} - v\left(\sqrt{a} - 1\right)^2 = 0, \tag{45}$$

This sum might be rewritten like integral $I = \frac{1}{2i} \oint_{C_1 \cup C_2} \frac{e^{i\pi z} dz}{\sin(\pi z)\left[b + \left(\pi z - \frac{\pi}{4}\right)^4\right]}$. Evaluating this integral using residue theory one can obtain [32]:

$$\frac{1}{b} + \sum_{n=1}^{\infty} \frac{1}{b + \left(\pi n - \frac{\pi}{4}\right)^4} = \frac{3}{4b} + \frac{1}{2\sqrt{2} b^{3/4}}, \tag{46}$$

where $v$ and $b$ are dimensionless coefficients:

$$v = \frac{K_f \cdot A_0}{\pi k_B T L^3}, \quad B = \frac{K_f}{L^4} \cdot b \tag{47}$$

So, the self-consistency equation takes the form:

$$\frac{3}{4b}+\frac{1}{2\sqrt{2}\,b^{3/4}}=v(\sqrt{a}-1)^2 \qquad (48)$$

We can solve equation (48) analytically and obtain expression for the entropic coefficient:

$$b=\frac{1}{4v^{4/3}(\sqrt{a}-1)^{8/3}}+\frac{1}{v(\sqrt{a}-1)^2} \qquad (49)$$

Then, the area per chain $a$ in flat monolayer may be found from equilibrium equation in flat bilayer [1,2]: $\frac{\partial F_t}{\partial A}=-\gamma$. The dependence $a(\gamma)$ is given in figure 10. Some calculation details are represented in App. E.

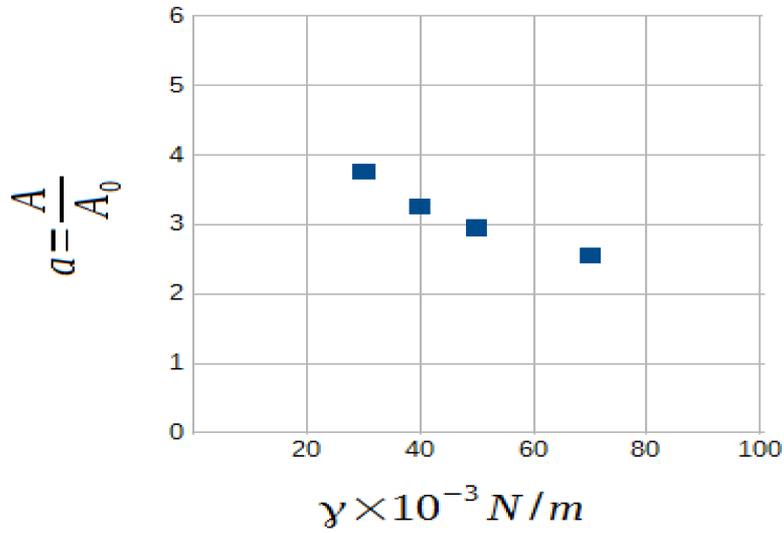

Figure 10 – The dependence of the equilibrium area per chain on surface tension between polar and hydrophobic regions $a(\gamma)$ in DPPC monolayer
( $L_0=1,5\,nm$, $T=300\,K$, $\gamma=50\,mN/m$ ).

### APPENDIX B: THE DERIVATIVES OF THE ENTROPIC COEFFICIENT

One can differentiate the self-consistency equation (48):

$$\left(\frac{3}{4b^2}+\frac{3}{8\sqrt{2}\,b^{7/4}}\right)\frac{\partial b}{\partial a}+v\frac{\sqrt{a}-1}{\sqrt{a}}=0 \qquad (50)$$

and find expression for the first derivative of entropic coefficient $\frac{\partial b}{\partial a}$:

$$\frac{\partial b}{\partial a}=-\frac{16v(\sqrt{a}-1)b^2}{3\sqrt{a}(4+\sqrt{2}\,b^{1/4})} \qquad (51)$$

Further, one can differentiate equation (50):

$$\left[\frac{3}{4b^2}+\frac{3}{8\sqrt{2}b^{7/4}}\right]\frac{\partial^2 b}{\partial a^2}-2\left[\frac{3}{4b^3}+\frac{21}{64\sqrt{2}b^{11/4}}\right]\left(\frac{\partial b}{\partial a}\right)^2+v\frac{1}{2a^{3/2}}=0 \quad (52)$$

and find expression for the second derivative $\frac{\partial^2 b}{\partial a^2}$:

$$\frac{\partial^2 b}{\partial a^2}=-\frac{8b^2 v}{3a^{3/2}\left(4+\sqrt{2}b^{1/4}\right)}+\frac{64}{9}\frac{v^2 b^3\left(\sqrt{a}-1\right)^2\left(32+7\sqrt{2}b^{1/4}\right)}{a\left(4+\sqrt{2}b^{1/4}\right)^3} \quad (53)$$

Some calculations results for derivatives of $b$ with respect to $a$ are represented in App. E

## APPENDIX C: THE LATERAL PRESSURE PROFILE AND MEMBRANE ELASTIC MODULI

The bending modulus are related to the first moment of the lateral pressure profile in bent monolayer [3, 33, 34]:

$$M_1=\kappa_m \cdot J=-\int_0^{L_0}\Pi_t^{(1)}(z,J)(z-z')dz;\quad \kappa_m=-\int_0^{L_0}\frac{\partial \Pi_t^{(1)}(z,J)}{\partial J}(z-z')dz, \quad (54)$$

where $L_0$ is the monolayer thickness and $z'$ indicates the neutral surface position. The so named first spontaneous bending moment is given in following form [34]:

$$M_{1s}=\kappa_m J_s=-\int_0^{L_0}(z-z')\Pi_t^{(0)}(z)dz \quad (55)$$

At the beginning we consider one monolayer of the lipid membrane, for example the 'bottom' (inner) monolayer ($z'=0$), and calculate the first lateral pressure moment and bending modulus (54). Substituting formula for the correction to lateral pressure profile for inner monolayer (33) into eq. (54) and integrating by $z$ one can obtain following expression:

$$\kappa_m J=kT\sum_n A\frac{\int_0^{L_0}\frac{\partial^2 B^{(0)}}{\partial A^2}\cdot z^2 R_n^{(0)2}(z)dz}{E_n^{(0)}}\cdot J-kT\sum_n A\frac{\left(\int_0^{L_0}\frac{\partial B^{(0)}}{\partial A}z R_n^{(0)2}(z)\right)^2}{E_n^{(0)2}}\cdot J, \quad (56)$$

where $A=A_0\cdot a$ - the area per lipid chain.

The we simplify (56) and re-write derivatives of $B$ and expression for eigen value (5) in dimensionless form using combinations $b=\frac{L_0^4}{K_f}\cdot B$ (App. B):

$$\kappa_m J=kT\sum_{n=0}^{\infty}\frac{a}{A_0}\frac{\int_0^{L_0}\frac{\partial^2 b^{(0)}}{\partial a^2}\cdot z^2 R_n^{(0)2}(z)dz}{b+c_n^4}\cdot J-kT\sum_{n=0}^{\infty}\frac{a}{A_0}\frac{\left(\int_0^{L_0}\frac{\partial b^{(0)}}{\partial a}z R_n^{(0)2}(z)\right)^2}{(b+c_n^4)^2}\cdot J, \quad (57)$$

Computing integrals in (57) one can obtain the following analytic expression for bending

modulus of monolayer:

$$\kappa_m = kT \frac{L_0^2}{A_0} a \frac{\partial^2 b}{\partial a^2} \left(\frac{1}{3b} + \sum_{n=1}^{\infty} \frac{I_2(n)}{b+c_n^4}\right) - kT \frac{L_0^2}{A_0} a \left(\frac{\partial b}{\partial a}\right)^2 \left(\frac{1}{4b^2} + \sum_{n=1}^{\infty} \frac{I_1^2(n)}{(b+c_n^4)^2}\right), \quad (58)$$

where expressions for the integrals are represented in the form ($c_n = \pi n - \frac{\pi}{4}$):

$$\int_0^{L_0} z R_n^{(0)2}(z) dz = L_0 \left(\frac{1}{2} + \frac{\sin(2c_n)}{2c_n} - \frac{\sin^2(c_n)}{2c_n^2}\right) = L_0 \cdot I_1(n) \quad (59)$$

$$\int_0^{L_0} z^2 R_n^{(0)2}(z) dz = \frac{L_0^2}{12}\left(4 + \frac{3\sin(2c_n)(2c_n^2-1)}{c_n^3}\right) = L_0^2 \cdot I_2(n) \quad (60)$$

Substitutung expression (16) into (55) one can calculate the spontaneous bending moment for inner monolayer $\kappa_m \cdot J_s$:

$$\kappa_m J_s = \int_0^{L_0} z \Pi_t^{(0)}(z) dz = \frac{kT}{A_0} \sum_{n=0}^{\infty} \frac{\int_0^{L_0} \frac{\partial b^{(0)}}{\partial a} z R_n^{(0)2}(z) dz}{b+c_n^4} \quad (61)$$

And after the substituting (59) into (61) one can obtain the analytic formula for the spontaneous bending moment:

$$\kappa_m \cdot J_s = \frac{kT}{A_0} L_0 \cdot \frac{\partial b^{(0)}}{\partial a}\left(\frac{1}{2b} + \sum_{n=1}^{\infty} \frac{I_1(n)}{b+c_n^4}\right) \quad (62)$$

The results (58), (62) are good agreement with previously obtained result for spontaneous bending moment $\kappa_m \cdot J_s = (-151 \pm 6) \times 10^{-13} J/m$ [25] and bending modulus of lipid monolayer [25], [34] (Table 3).

| $L_0, nm$ | $\kappa_m, k_B T$ | $\kappa_m J_s \times 10^{-12} J/m$ |
| --- | --- | --- |
| 1 | 9,24 | -19,12 |
| 1,2 | 13,3 | -22,88 |
| 1,5 | 20,6 | -28,6 |
| 2,0 | 35,7 | -38,24 |

Table 3 – The spontaneous bending moment and the bending modulus for lipid monolayer calculated analytically in the microscopic flexible string model of the lipid monolayer. The input

parameters of lipid chain: $A_0 = 0,2\,nm^2$ incompressible area, $T = 300\,K$ temperature, $\gamma = 50\,mN/m$ is the surface tension at the interface between the polar and hydrophobic regions of the membrane. Here $L_0$ - the monolayer thickness. The dependence $a(\gamma, L_0)$ is found from equilibrium equation for flat bilayer [1, 2] $\frac{\partial F_t}{\partial A} = -\gamma$ - eq.(14).

The analyze of bilayer bending/stretching moduli values is represented in table 4. Usually experimental value for bending modulus for different lipid bilayers vary between $(0,1-6) \times 10^{-19} J$ [36], [37], [38].

The relation between bilayer bending modulus and area stretching modulus of bilayer obtained from our calculation is: $\kappa_b \approx \frac{1}{12} \cdot K_{Ab} L_b^2$, where $L_b = 2 L_0$ - the bilayer thickness and $\kappa_b = 2 \kappa_m$ the bending rigidity for bilayer (table 4). This condition is in agreement with [38, 39]. and considered as correct one for symmetrical monolayers.

| $L_0$, nm | $\gamma = 50, mN/m$ | | $\gamma = 40, mN/m$ | | $\gamma = 30, mN/m$ | |
|---|---|---|---|---|---|---|
| | $\kappa_b, k_B T$ | $K_A, N/m$ | $\kappa_b, k_B T$ | $K_A, N/m$ | $\kappa_b, k_B T$ | $K_A, N/m$ |
| 1 | 18,5 | 0,2336 | 13,169 | 0,17169 | 9,34 | 0,12 |
| 1,2 | 26,6 | 0,23455 | 19,71 | 0,17405 | 13,5 | 0,12 |
| 1,5 | 41,2 | 0,2326 | 30,8 | 0,174 | 21,173 | 0,12 |
| 2,0 | 71,5 | 0,226 | 54 | 0,172113 | 37,65 | 0,12 |

Table 4 – The calculation of the bending modulus for various bilayer and its comparison with stretching modulus in flexible string model; $L_0$ - the monolayer thickness, $\kappa_b$ - the bending modulus of the bilayer, stretching modulus for the bilayer $K_{Ab} = 2 A \cdot \frac{\partial^2 F_t}{\partial A^2}$. Itsthe expression in flexible strings models is: $K_{Ab} = 2 a \frac{kT}{A_0} (\frac{\partial^2 b}{\partial a^2} \sum_{n=0}^{\infty} \frac{1}{b + c_n^4} - (\frac{\partial b}{\partial a})^2 \sum_{n=0}^{\infty} \frac{1}{(b + c_n^4)^2})$.

**APPENDIX D: THE CALCULATION OF INTEGRALS CONSISTED IN THE ENERGY DIFFERENCE BETWEEN TWO CONFORMATIONAL STATES OF CHANNEL**

The integral (55) for conical shape of the channel $\int_0^{2L_0} \Pi_t^{(0)}(z) \Delta A(z) dz$ (fig. 4 B) can be calculated as follows (eq. (37)):

$$\int_0^{2L_0} \Pi_t^{(0)}(z)\Delta A(z)dz \sim$$

$$\sim \alpha A_m \int_0^{L_0} z R_n^{(0)2}(z)dz - \alpha A_m L_0 \int_0^{L_0} R_n^{(0)2}(z)dz - \qquad (63)$$

$$-\alpha A_m \int_0^{L_0} z R_n^{(0)2}(z)dz + \alpha A_m L_0 \int_0^{L_0} R_n^{(0)2}(z)dz \equiv 0$$

The same integral in hour-glass model (fig. 4C) (eq. (38), (55), (62)) takes the form:

$$\int_0^{2L_0} \Pi_t^{(0)}(z)\Delta A(z)dz \sim$$

$$\sim -2\alpha A_m \int_0^{L_0} z R_n^{(0)2}(z)dz + 2\alpha A_m L_0 \int_0^{L_0} R_n^{(0)2}(z)dz \sim 2\alpha A_m \kappa_m J_s + 2\alpha A_m L_0 = \qquad (64)$$

$$= 2\alpha A_m \kappa_m J_s + 2\Delta A_{max}\gamma$$

The same integral in hour-glass model (fig. 4D) (eq. (39), (55), (62)) can be calculated similarly:

$$\int_0^{2L_0} \Pi_t^{(0)}(z)\Delta A(z)dz \sim$$

$$\sim 2\alpha A_m \int_0^{L_0} z R_n^{(0)2}(z)dz - 2\alpha A_m L_0 \int_0^{L_0} R_n^{(0)2}(z)dz \sim -2\alpha A_m \kappa_m J_s - 2\alpha A_m L_0 = \qquad (65)$$

$$= -2\alpha A_m \kappa_m J_s - 2\Delta A_{max}\gamma$$

The integral $\int_0^{2L_0} \Pi_t^{(1)}(z)\Delta A(z)dz$ in conical model (fig. 4B) (eq. (33), (34), (37), (54)) is equiv to the integrals:

$$\int_0^{2L_0} \Pi_t^{(1)}(z)\Delta A(z)dz \sim$$

$$\sim -\int_0^{L_0} \alpha A_m(z-L_0)z R_n^{(0)2}(z)dz + \int_0^{L_0}(-\alpha A_m(z-L_0))z R_n^{(0)2}(z)dz \qquad (66)$$

$$= -2\alpha A_m \int_0^{L_0} z^2 R_n^{(0)2}(z)dz + 2\alpha A_m L_0 \int_0^{L_0} z R_n^{(0)2}(z)dz \sim$$

$$= -2\alpha A_m \kappa_m J + \Delta A_{max}\tilde{K}_{Ab} L_0 J$$

where expression for re-normalized stretching modulus one can view in sec. V. Here for clarity the integral $(\int_0^{L_0} z R_n^{(0)2}(z)dz)^2$ in (33)-(34) wasn't mentioned, but was keep in mind.

The same integral in hour-glass model (fig. 4C):

$$\alpha A_m \int_0^{L_0}(z-L_0)z R_n^{(0)2}(z)dz - \alpha A_m \int_0^{L_0}(z-L_0)z R_n^{(0)2}(z)dz \equiv 0 \qquad (67)$$

The calculations $A_m \int_0^{2L_0} \Pi_t^{(1)}(z)\,dz$ gives zero result in the first-order approximation (eq. (33)-(34)).

## APPENDIX E: CALCULATIONS DETAILS

In this section we present the solution of the self-consistency equation (48) for various lipids and also the derivatives of the entropic coefficient (51), (53). The input parameters are: incompressible area per chain $A_0 = 0{,}2\,nm^2$, dimensionless parameter $\nu = \dfrac{A_0 K_f}{\pi kT L^3}$, and the bending rigidity of the lipid chain $K_f = kTL_0/3$.

| $\gamma/L_0$ | 1,0 | 1,2 | 1,5 | 2,0 | 2,2 | 2,4 | 2,5 |
|---|---|---|---|---|---|---|---|
| 30 | 3,49 | 3,6 | 3,76 | 4,01 | 4,103 | 4,19 | 4,232 |
| 40 | 3,02 | 3,12 | 3,27 | 3,48 | 3,555 | 3,627 | 3,662 |
| 50 | 2,73 | 2,82 | 2,95 | 3,13 | 3,2 | 3,262 | 3,2915 |
| 70 | 2,37 | 2,45 | 2,56 | 2,71 | 2,757 | 2,807 | 2,831 |

Table 5 — The calculation of the area per lipid chain $a$, $\gamma$, $mN/m$, $L_0$, $nm$

| $\gamma/L_0$ | 1,0 | 1,2 | 1,5 | 2,0 | 2,2 | 2,4 | 2,5 |
|---|---|---|---|---|---|---|---|
| 30 | 124,56 | 176,87 | 268,6 | 455,69 | 543,53 | 636,4 | 685,9 |
| 40 | 182,3 | 256 | 385,14 | 649,3 | 772,8 | 906 | 976,2 |
| 50 | 244,81 | 341,12 | 510,4 | 859,5 | 1022,3 | 1198,4 | 1291,6 |
| 70 | 381,45 | 528,46 | 784,4 | 1324,6 | 1576,19 | 1846,2 | 1988,6 |

Table 6 — The calculation of the entropic coefficient $b$, $\gamma$, $mN/m$, $L_0$, $nm$

| $\gamma/L_0$ | 1,0 | 1,2 | 1,5 | 2,0 | 2,2 | 2,4 | 2,5 |
|---|---|---|---|---|---|---|---|
| 30 | -92,28 | -124,3 | -177,2 | -276,5 | -319,7 | -365 | -388,3 |
| 40 | -170 | -226,9 | -320,3 | -494,82 | -571,4 | -651,1 | -692 |
| 50 | -273,2 | -361,4 | -506,5 | -779,6 | -898,4 | -1022,8 | -1087,33 |
| 70 | -555,9 | -730,2 | -1015 | -1554 | -1790,8 | -2035,3 | -2161,4 |

Table 7 — The calculation of the first derivative of the entropic coefficient $\dfrac{\partial b}{\partial a}$, $\gamma$, $mN/m$, $L_0$, $nm$

| $\gamma/L_0$ | 1,0 | 1,2 | 1,5 | 2,0 | 2,2 | 2,4 | 2,5 |
|---|---|---|---|---|---|---|---|
| 30 | 109,9 | 142,5 | 193 | 278,3 | 312,4 | 346,7 | 363,8 |
| 40 | 255,6 | 327,6 | 436,6 | 616,8 | 689,7 | 762,6 | 799 |
| 50 | 491 | 620,8 | 815,9 | 1142 | 1272,9 | 1045,5 | 1473 |
| 70 | 1295,6 | 1613,7 | 2085 | 2891 | 3224,3 | 3555,12 | 3722,3 |

Table 8 — The calculation of the second derivative of the entropic coefficient $\frac{\partial^2 b}{\partial a^2}$, $\gamma, mN/m$, $L_0, nm$

**APPENDIX F: THE CALCULATION OF ENERGY DIFFERENCE BETWEEN TWO DIFFERENT CONFORMATIONAL STATES OF CHANNEL**

In this section we represent the calculated energy difference between two states of mechanosensitive channels for different lipids. We change the monolayer thickness and the polar tension as the input parameters. As one can see from table 9 the absolute value of work associated with embedding the channel in its hour-glass conformation into flat membrane ($\Delta W_{A \to C} / \Delta W_{A \to D}$) does not depend on the monolayer thickness. This is caused by the fact that spontaneous bending energy $|2\alpha A_m \kappa_m J_s|$ decreases/increases the hydrophobic energy $|2\Delta A_{max} \gamma|$ to permanent value. Therefore for all thicknesses of the lipid monolayer we obtain the almost constant energy difference depended majorly on the renormalized inter-facial surface tension $\Delta W_{A \to C} \approx \beta \Delta A_{max} \cdot \gamma$, where $\beta = 1,235 \pm 0,005$ - the renormalization constant (the value of constatnt $\beta$ may be obtained after substituting eq. (62) and expression for inter-facial tension ($\frac{\partial F}{\partial A} = -\gamma$) into formula for energy difference $\Delta W_{A \to C}$). This result is well consistent with previous calculation [23].

One can note that expressions for:
- energy difference for transition from shape $A$ to shape $B$ is represented in the form (f. (30), (37) and (66)) :

$$\Delta W_{A \to B} = -2\alpha A_m \kappa_m J + \Delta A_{max} \tilde{K}_{Ab} L_0 J \qquad (68)$$

- energy difference for transition from shape $A$ to shape $C$ is given by formula (f. (16), (38) and (64)) :

$$\Delta W_{A \to C} = 2\alpha A_m \kappa_m J_s + 2\Delta A_{max} \gamma \qquad (69)$$

- energy difference for transition from shape $B$ to shape $C$ is given by equation (f. (16), (30), (33), (37), (38), (64), (66)):

$$\Delta W_{B \to C} = 2\alpha A_m \kappa_m J_s + 2\Delta A_{max} \gamma + 2\alpha A_m \kappa_m J - \Delta A_{max} \tilde{K}_{Ab} L_0 J \qquad (70)$$

- energy difference for transition from shape $B$ to shape $D$ may be obtained similarly (f. (16), (30), (34), (37), (38), (64), (66)):

$$\Delta W_{B \to D} = -2\alpha A_m \kappa_m J_s - 2\Delta A_{max} \gamma + 2\alpha A_m \kappa_m J - \Delta A_{max} \tilde{K}_{Ab} L_0 J \qquad (71)$$

Finally, we can conclude that bending of membrane associated with simple cylinder-cone-like shape channel transition doesn't produce a great changing in energy difference. Hydrophobic energy reduced the contact between polar heads and non-polar tails is the primary stimulus for MSC gating.

It is important to note that calculation performed in this work may be applied to any protein (not only mechanosensitive channel) which undergoes similar conformational changes.

| $\Delta W$, $k_B T$ | | $L_0$, nm | | | | | | |
|---|---|---|---|---|---|---|---|---|
| | $\gamma$, mN/m | 1,0 | 1,2 | 1,5 | 2,0 | 2,2 | 2,4 | 2,5 |
| $\Delta W_{A \to B}$ | 30 | 0,19 | 0,24 | 0,34 | 0,49 | 0,55 | 0,61 | 0,64 |
| | 40 | 0,28 | 0,37 | 0,5 | 0,71 | 0,79 | 0,86 | 0,9 |
| | 50 | 0,39 | 0,51 | 0,68 | 0,93 | 1,02 | 1,12 | 1,16 |
| $\Delta W_{A \to C}$ | 30 | 13,08 | 13,14 | 13,18 | 13,2 | 13,2 | 13,19 | 13,19 |
| | 40 | 17,53 | 17,57 | 17,59 | 17,59 | 17,57 | 17,56 | 17,56 |
| | 50 | 21,96 | 21,99 | 22 | 21,96 | 21,94 | 21,93 | 21,92 |
| $\Delta W_{B \to C}$ | 30 | 12,89 | 12,89 | 12,84 | 12,71 | 12,64 | 12,58 | 12,55 |
| | 40 | 17,25 | 17,2 | 17,09 | 16,88 | 16,79 | 16,7 | 16,66 |
| | 50 | 21,56 | 21,48 | 21,31 | 21,03 | 20,92 | 20,81 | 20,75 |
| $\Delta W_{B \to D}$ | 30 | -13,27 | -13,38 | -13,52 | -13,69 | -13,75 | -13,8 | -13,83 |
| | 40 | -17,81 | -17,94 | -18,09 | -18,3 | -18,36 | -18,42 | -18,46 |
| | 50 | -22,35 | -22,5 | -22,67 | -22,89 | -22,97 | -23,04 | -23,08 |

Table 9 – The calculated energy difference between two conformational shape of a mechanosensitive channel for different transition and various lipids thicknesses. Here $\Delta W_{A \to B}$ - the energy difference for channel transition from shape $A$ to shape $B$ and other energy differences may be defined similarly. $L_0$ - the monolayer thickness, $\gamma$ - interfacial surface tension. One can note that transitions between channel shapes are characterized by relations:

$\Delta W_{A \to B \to C} = \Delta W_{A \to B} + \Delta W_{B \to C} \equiv \Delta W_{A \to C}$, $\qquad \Delta W_{A \to D} = -\Delta W_{A \to C}$,

$\Delta W_{A \to B \to D} = \Delta W_{A \to B} + \Delta W_{B \to D} = -\Delta W_{A \to B \to C} \equiv -\Delta W_{A \to C}$. For calculation the first bending moment was taken the curvature radius $R = 20\, nm$ and curvature for spherical surface $J = \dfrac{2}{R}$.